\documentclass[a4paper,11pt]{article}
\pdfoutput=1 
\usepackage{jcappub} 

\usepackage[T1]{fontenc} 
\usepackage{graphicx}
\usepackage{longtable}
\usepackage{float}
\usepackage{dcolumn}
\usepackage{bm}
\usepackage{appendix}
\usepackage{multirow}
\usepackage{color}

\newcommand{\Mpc}{\mathrm{Mpc}}

\title{\boldmath Testing non-minimally coupled inflation with CMB data: a Bayesian analysis}

\author[a,b]{Marcela Campista}

\author[a]{Micol Benetti}

\author[a]{Jailson Alcaniz}

\affiliation[a]{Observat\'orio Nacional, Rua General Jos\'e Cristino 77,  20921-400, Rio de Janeiro, RJ, Brazil}
\affiliation[b]{Instituto de F\'isica, Universidade Federal do Rio de Janeiro, 21941-972, Rio de Janeiro, RJ, Brazil}

\emailAdd{campista@on.br}
\emailAdd{micolbenetti@on.br}
\emailAdd{alcaniz@on.br}
\date{\today}

\abstract{
We use the most recent cosmic microwave background (CMB) data to perform a Bayesian statistical analysis and discuss the observational viability of inflationary models with a non-minimal coupling,~$\xi$, between the inflaton field and the Ricci scalar. We particularize our analysis to two examples of small and large field inflationary models, namely, the Coleman-Weinberg and the chaotic quartic potentials. We find that (\textit{i}) 
the $\xi$ parameter is closely correlated with the primordial amplitude
; (\textit{ii}) although improving the agreement with the CMB data in the $r - n_s$ plane, where $r$ is the tensor-to-scalar ratio and $n_s$  the primordial spectral index, a non-null coupling is strongly disfavoured with respect to the minimally coupled standard $\Lambda$CDM model, since the upper bounds of the Bayes factor (odds) for $\xi$ parameter are greater than $150:1$.
}

\begin{document}
\maketitle
\flushbottom

\section{Introduction}
\label{Introduction}

The observational viability of a wide range of inflationary models has now been tested thanks to the accuracy of current Cosmic Microwave Background (CMB) data~\cite{Ade:2015xua, Ade:2015lrj}. The observations are compatible with both the simplest slow-roll scenarios of inflation, showing a preference for plateau over monomial potentials, as well as with some alternative scenarios (see, e.g., Refs.~\cite{Ijjas:2013vea,Linde:2014nna,Guth:2013sya,Brandenberger:2015kga,Martin:2015dha,Benetti:2016tvm} for different points of view of the current observational status of inflation). Since no compelling statistical evidence has been found for a specific inflationary model, an important task nowadays is to examine the theoretical predictions of different classes of scenarios in the light of current data. 

Among these scenarios, models in which the scalar field is taken to be non-minimally coupled to the scalar curvature have been widely discussed in the literature~(see, e.g. \cite{Linde:2011nh} and references therein). Recently,  a broad class of non-minimally coupled chaotic inflation with arbitrary potential has  been proposed in the context of supergravity \cite{Kallosh:2010ug} as well as the possibility of implementing chaotic inflation  with the Higgs field non-minimally coupled to gravity playing the role of the inflaton field~\cite{Bezrukov:2007ep,Barvinsky:2008ia,GarciaBellido:2008ab,DeSimone:2008ei}. From the observational point of view, it has been shown that the introduction of a non-minimal coupling to gravity may improve the description of the data (e.g., predicting reasonable amplitude of the density perturbations) without fine-tuning the self-coupling constant of the inflaton field~\cite{Fakir:1990eg,Komatsu:1997hv,Komatsu:1999mt,Bezrukov:2008dt,Linde:2011nh}. In this context, single and multifield models of inflation have also been explored~\cite{Kaiser:2013sna, Chen:2014zoa} and, across a broad range of couplings and initial conditions, such models evolve along an effectively single-field attractor solution and may predict values of the primordial spectral index, $n_s$, and of tensor-to-scalar ratio, $r$, in the observationally most-favoured region with respect to the minimal coupling case~\cite{Okada:2010jf, Okada:2015lia, Bezrukov:2008dt, Linde:2011nh, Komatsu:1999mt}.

Differently from previous studies, here we go beyond the analysis of compatibility in the $r - n_s$ plane and perform a Bayesian statistical analysis to investigate whether the most recent Planck release~\cite{Ade:2015xua} support the idea of a non-minimally coupled inflation when compared with the $\Lambda$CDM model. We particularize our analysis to two examples of small field and large field inflationary potentials, namely, the Coleman-Weinberg~\cite{Coleman:1973jx} and the chaotic quartic~\cite{Fakir:1990eg,Komatsu:1997hv,Komatsu:1999mt,Bezrukov:2008dt} potentials.  We found that, although improving the agreement in the $r - n_s$ plane, a non-null coupling is not enough to attest the observational viability of these models and both are strongly disfavoured by the current CMB data with respect to the standard scenario.

We organized this paper as follows. Sec.~\ref{Frame} reviews the non-minimal coupling formalism. In Sec.~\ref{Model} we introduce the Coleman-Weinberg and the chaotic potentials with a non-minimal  coupling to gravity.  In  Sec.~\ref{Method} we discuss our analysis approach and the observational data sets used in the analysis. In Sec.~\ref{Results} we present a brief comparison with previous analyses and summarize in Sec.~\ref{Conclusions} our main results.

\section{Non-Minimal Coupling and Inflation} 
\label{Frame}

In the inflationary context, the idea of non-minimal coupling requires the replacement of the Ricci scalar by the term $\xi\varphi^2R$ in the inflationary action:
\begin{equation} \label{NM_action}
 S=\int{d^4 x \sqrt{-{g}}\left[\frac{1+\kappa ^2 \xi \varphi^2 }{2 \kappa^2}R -\frac{1}{2}{g}_{\mu\nu}\partial\varphi_{\mu}\partial\varphi_{\nu}- U(\varphi)\right]},
\end{equation}
where $\kappa=8\pi G$ and $\xi$ is the coupling constant, responsible for the length of interaction. Differently from  other theories where the coupling $\xi$ is fixed by the theory of scalar field assumed \cite{Faraoni:1997fn}, here 
it is treated as a free parameter and constrained by the observable data. The minimal inflationary scenario is recovered by setting
the coupling parameter $\xi=0$. 

The inclusion of a coupling in the Lagrangian makes the field equations more complicated, and the standard procedure is 
to consider a conformal transformation in the metric tensor $g_{\mu\nu}\rightarrow \Omega^{2} g_{\mu\nu}$, with
$\Omega^2= 1+ \kappa^2\xi \varphi^2$, so that the coupling constant is eliminated in the gravitational sector and introduced
inside a new potential by the relation $V(\phi) = \frac{U(\varphi(\phi))}{\Omega^{4}}$. The frame in which the non-minimal coupling
is eliminated by conformal transformation is called Einstein frame~\cite{Okada:2010jf} and the new action is written as
\begin{equation}
 S=\int{d^4 x \sqrt{-{g}}\left[R -\frac{1}{2}{g}_{\mu\nu}\partial\phi_{\mu}\partial\phi_{\nu}- V(\phi)\right]}.
 \label{eq:non-minimal_action}
\end{equation}
It is important to note that, in this frame, a canonical kinetic term arises with the new inflaton field $\phi$ related to the original 
inflaton field $\varphi$ by
\begin{equation}\label{fields}
\frac{d\phi}{d\varphi} = \frac{\sqrt{1+ \kappa ^2 \xi \varphi^{2} (1+6\xi)}}{\Omega^{2}}.
\end{equation}

Independently of frame, the inflationary dynamics is governed by the slow-roll parameters, $\epsilon$ and $\eta$~\cite{Mukhanov}. However, the inclusion of the non-minimal coupling modifies the potential in 
the Einstein frame and, for this reason, the slow-roll parameters need to be redefined:
\begin{subequations}
\label{inflation_params}
\begin{equation}
\epsilon = \frac{1}{2}\left(\frac{d\varphi}{d\phi}\right)^2\left(\frac{V^{\prime}}{V}\right)^2  ,
\end{equation}
\begin{equation}
\eta \equiv\frac{V^{\prime\prime}}{V}\left(\frac{d\varphi}{d\phi}\right)^{2}-\left(\frac{d^2\phi}{d\varphi^2}\right)\frac{V^{\prime}}{V}\left(\frac{d\varphi}{d\phi}\right)^{3},
\end{equation}
\end{subequations}
where a prime denotes the derivative with respect to $\varphi$. A successful slow-roll inflation occurs while the slow-roll parameters are small, $\epsilon << 1$ and  $\eta << 1$. When this condition is violated, e.g. $\epsilon \simeq 1$, the potential reaches its minimum and the inflation period ends,
with $\phi=\phi_{end}$.

The power spectrum of the curvature perturbation is given by
\begin{equation}
P_R =\frac{V(\phi)}{24\pi^2\epsilon}|_{k=k_{*}}\;,
\label{eq:PR}
\end{equation}
where $(\ast)$ refers to 
the (pivot) scale at which the CMB mode exits from horizon. 
Also, the value of $P_{R}(k_{\ast})$ is established by the COBE normalization, i.e., $\sim2.2 \times 10^{-9}$ for the pivot choice $k_{*}=0.05 {\rm{Mpc^{-1}}}$ \cite{Ade:2015xua}. As usual, the standard inflationary parameters, i.e., the spectral index and tensor-to-scalar ratio, are defined in terms of slow-roll parameters calculated in the horizon crossing
\begin{equation} \label{eq:parameters}
 n_s=-6\epsilon+2\eta+1 
\qquad \rm{and} \qquad 
 r=16\epsilon .
\end{equation}
Throughout this paper, we shall work in reduced Planck units ($c = h = 8\pi G = 1$).

\begin{figure} []
\centering
\includegraphics[height=0.7\hsize, angle=-90]{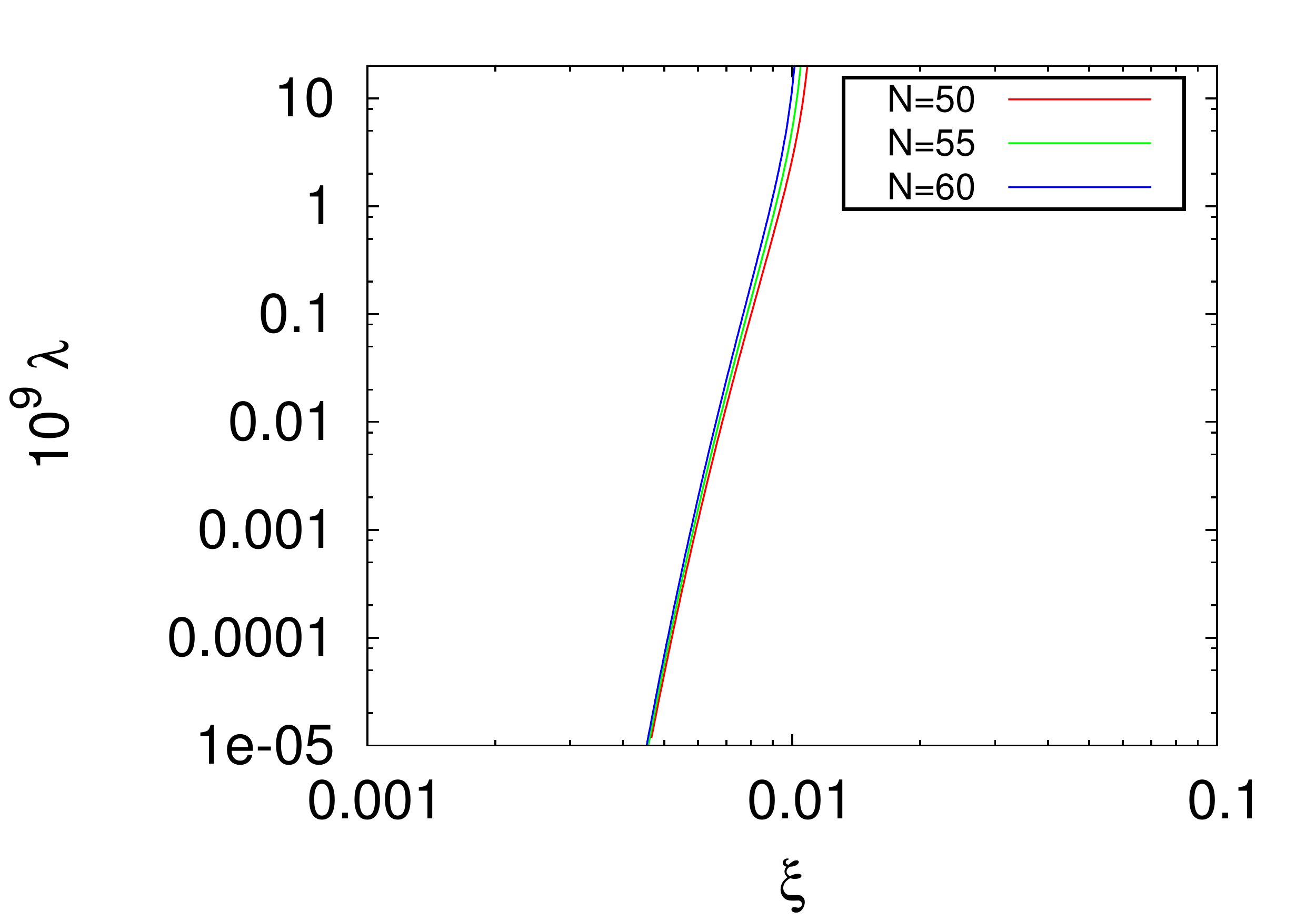}
\caption{The behaviour of the amplitude $\lambda$ for the CW potential as a function of the non-minimal coupling parameter $\xi$ for different values of number of e-folds $N$. \label{fig:CW_lambda_xi}}
\end{figure}
\section{Inflationary Models}
\label{Model}

In what follows, we study the observational predictions of two examples of non-minimally coupled small field and large field inflation, namely, the Coleman-Weinberg~\cite{Coleman:1973jx} and the chaotic quartic~\cite{Fakir:1990eg,Komatsu:1997hv,Komatsu:1999mt,Bezrukov:2008dt}.
\begin{figure*} []%
\centering
\includegraphics[height=0.4\hsize]{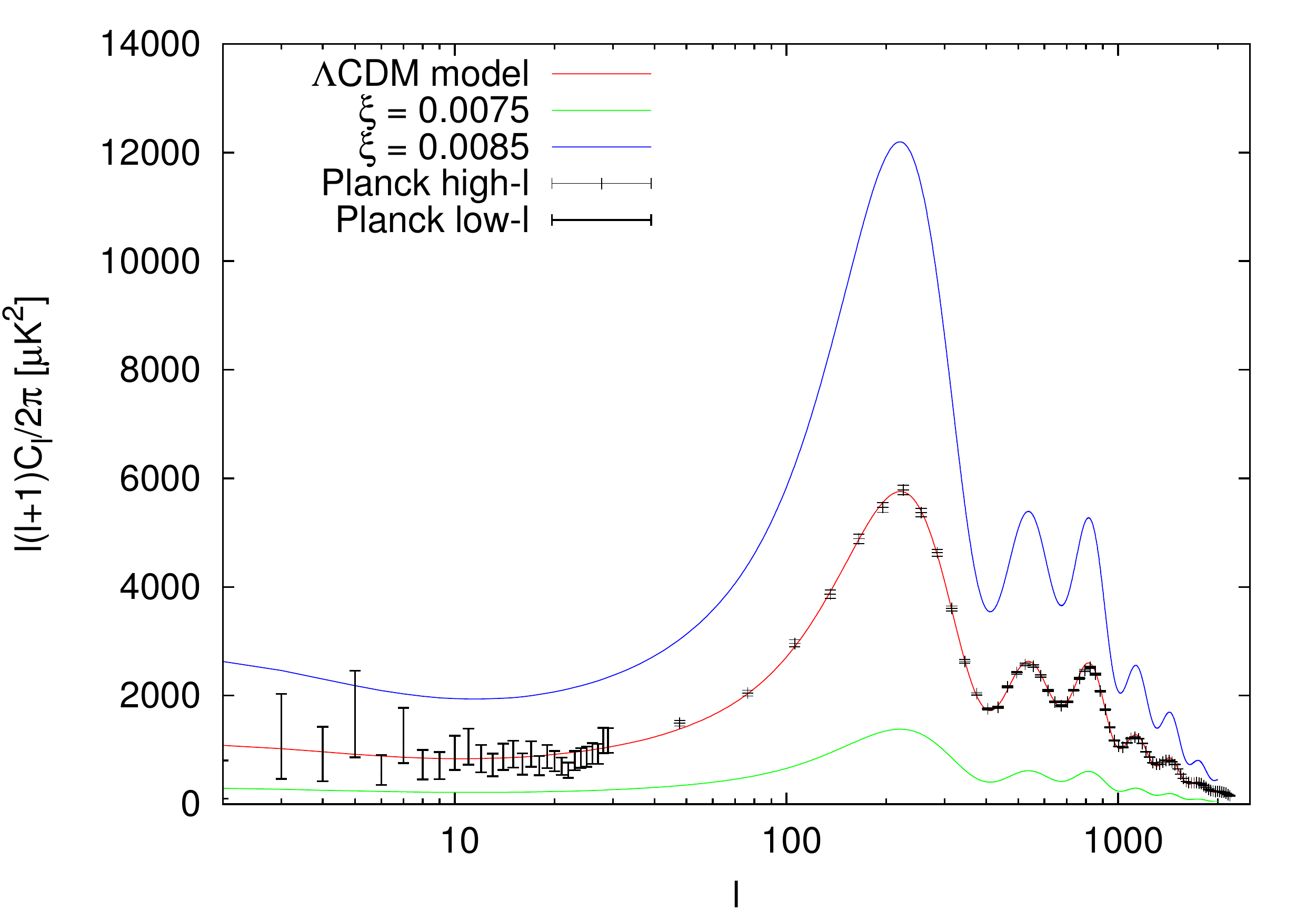}
\includegraphics[height=0.4\hsize]{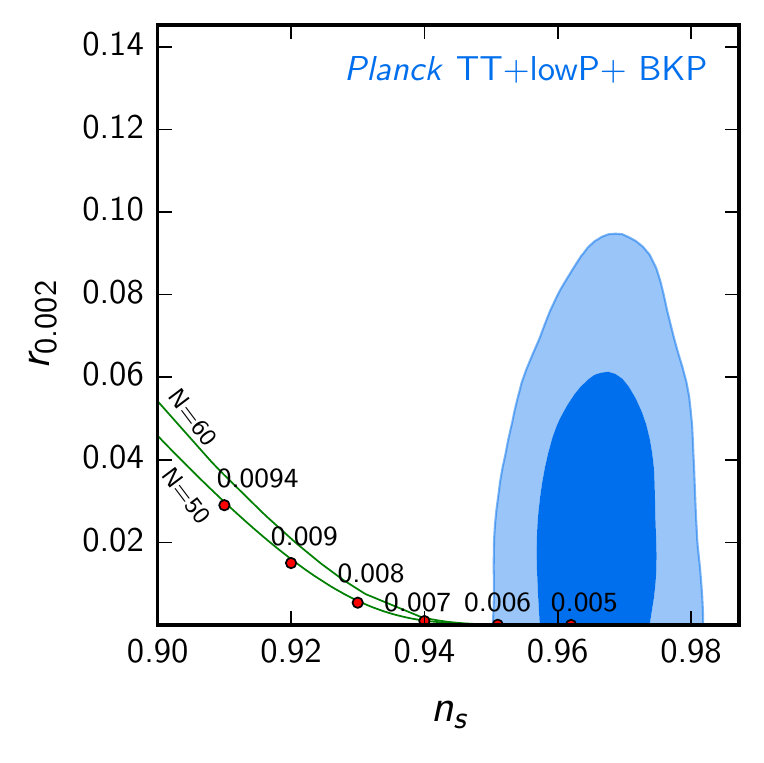}
\caption{{\textit{Left}}: The CMB power spectrum prediction of non-minimally coupled CW models for selected values of the coupling $\xi$, compared with the most recent Planck data. {\it{Right: }} The $n_s - r$ plane. The contours correspond to $68 \%$ and $95\%$ regions obtained from the Plank TT+lowP data combined with the  Bicep$/$Keck Array data~\cite{Ade:2015tva}. The green lines correspond to the evolution of the $n_s$ and $r$ values for increasing values of $\xi$ (red dots) assuming $N=50$ and $N=60$.} \label{fig:CW_predictions}
\end{figure*}

\subsection{Coleman-Weinberg (CW) potential} 

The Coleman-Weinberg (CW) potential, introduced in Ref.~\cite{Coleman:1973jx}, 
can be written as:

\begin{equation}\label{eq: CW_potential_minimal}
V(\varphi)=V_{o}\left[1-\frac{\varphi^4}{M^4}+4\frac{\varphi^4}{M^4}\log(\frac{\varphi}{M})\right]\;,
\end{equation}
where $M$ is the inflaton vacuum expectation value $\langle \phi \rangle$ at  minimum potential. 

The minimal coupling version of this model was recently analysed in \cite{Barenboim:2013wra}, where it was shown that the predicted spectral index lies outside the $2\sigma$ region provided by the CMB Planck data. For the non-minimal coupling CW model, the solution of Eq. (\ref{fields}) in the small
field regime provides $\varphi=\phi$, which means that Eq. (\ref{eq: CW_potential_minimal}) can be rewritten as
\begin{equation}
V(\phi)=\lambda(1-2\kappa^2 \xi\phi^2).
\label{eq:Potential_CW}
\end{equation}
The slow-roll parameters for the potential above can be easily derived from Eq. (\ref{inflation_params}), i.e.,
\begin{equation} \label{epsilon}
 \epsilon = \frac{8\xi^2\phi^2}{(1-2\xi\phi^2)^2}\;,
\qquad
\eta =-\frac{4V_{o}\xi}{(1-2\xi\phi^2)^2}\;,
\end{equation}
and the condition $\epsilon_{end} \simeq 1$ implies 
\begin{equation}
 \phi_{end}= \pm \frac{\sqrt{2(1+4\xi)\xi}}{2\xi}.
\end{equation}
Using the COBE normalization \cite{Ade:2015xua}, we can easily write the dependence of amplitude 
$\lambda$ with $\xi$ at the horizon exit as
\begin{equation} \label{lcw}
 \lambda=\frac{192\pi^2\xi^2{\phi_{\ast}}^2 P_{R}(k_{\ast})}{(1-2\xi{\phi_{\ast}}^2)^3}.
\end{equation}
In Fig.~(\ref{fig:CW_lambda_xi}) we show the  behaviour of $\lambda$ as a function of $\xi$, noting that the number of e-folds, defined as $N=\int^{\phi_{\ast}}_{\phi_{end}}{d\phi}/{\sqrt{2\epsilon}}$, does not significantly influence the result. Thus, hereafter we assume $N = 50$ for this model. The value of the scale at the horizon crossing, $\phi_*$, can be obtained from the above equations, i.e.,
 \begin{equation}
N = \frac{1}{4\xi}\left[\log\left(\frac{\phi_{*}}{\phi_{end}}\right)\right]  - \frac{\phi_{*}^2}{4}+\frac{\phi_{end}^2}{4} .
\label{eq:efold_CW}
 \end{equation}
By solving numerically the above equation we find no real solutions of $\phi_{\ast}$ for values of $\xi$ higher than  $0.012$. 
\begin{figure*}[] %
\centering
\includegraphics[height=0.35\hsize]{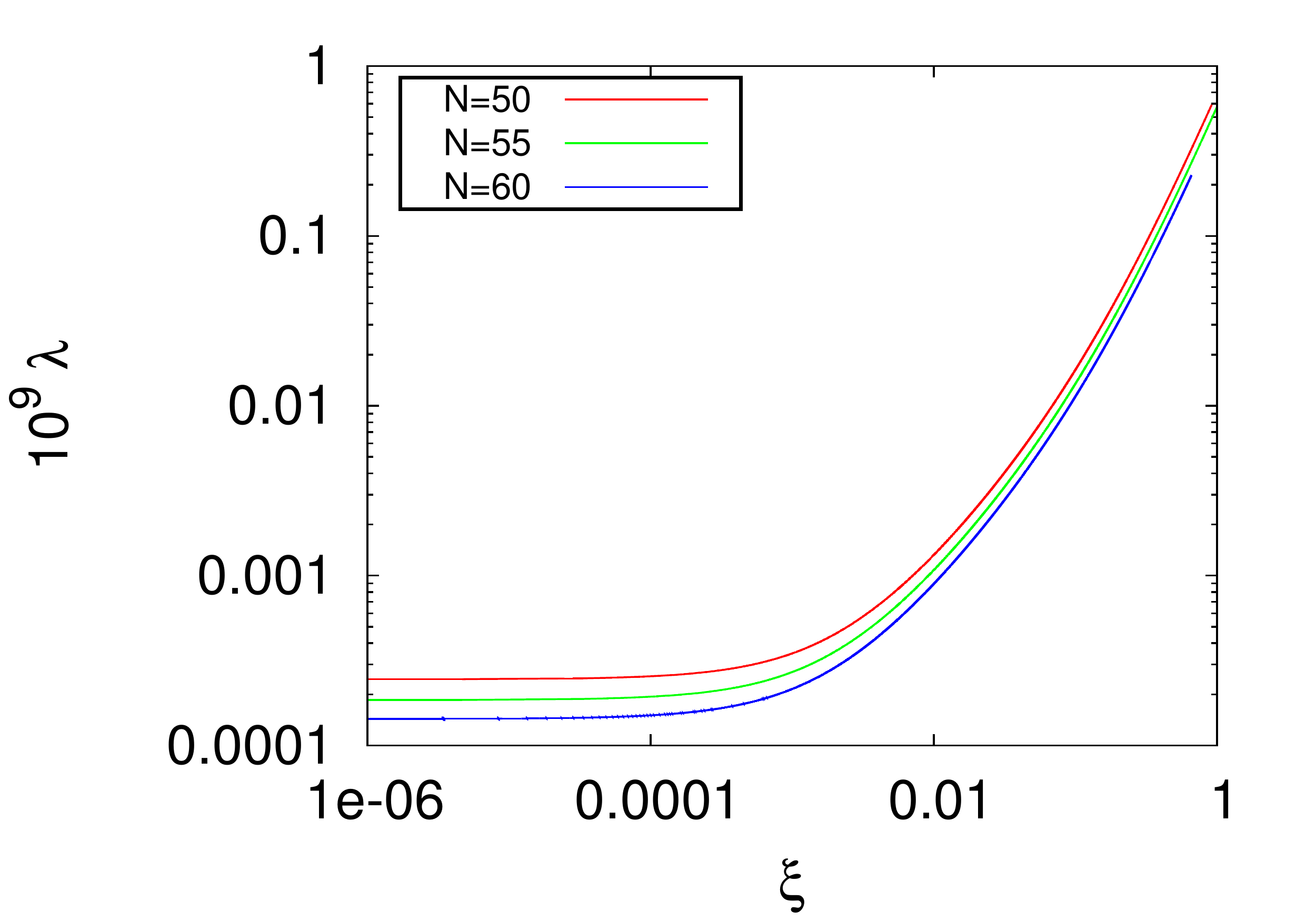}
\includegraphics[height=0.35\hsize]{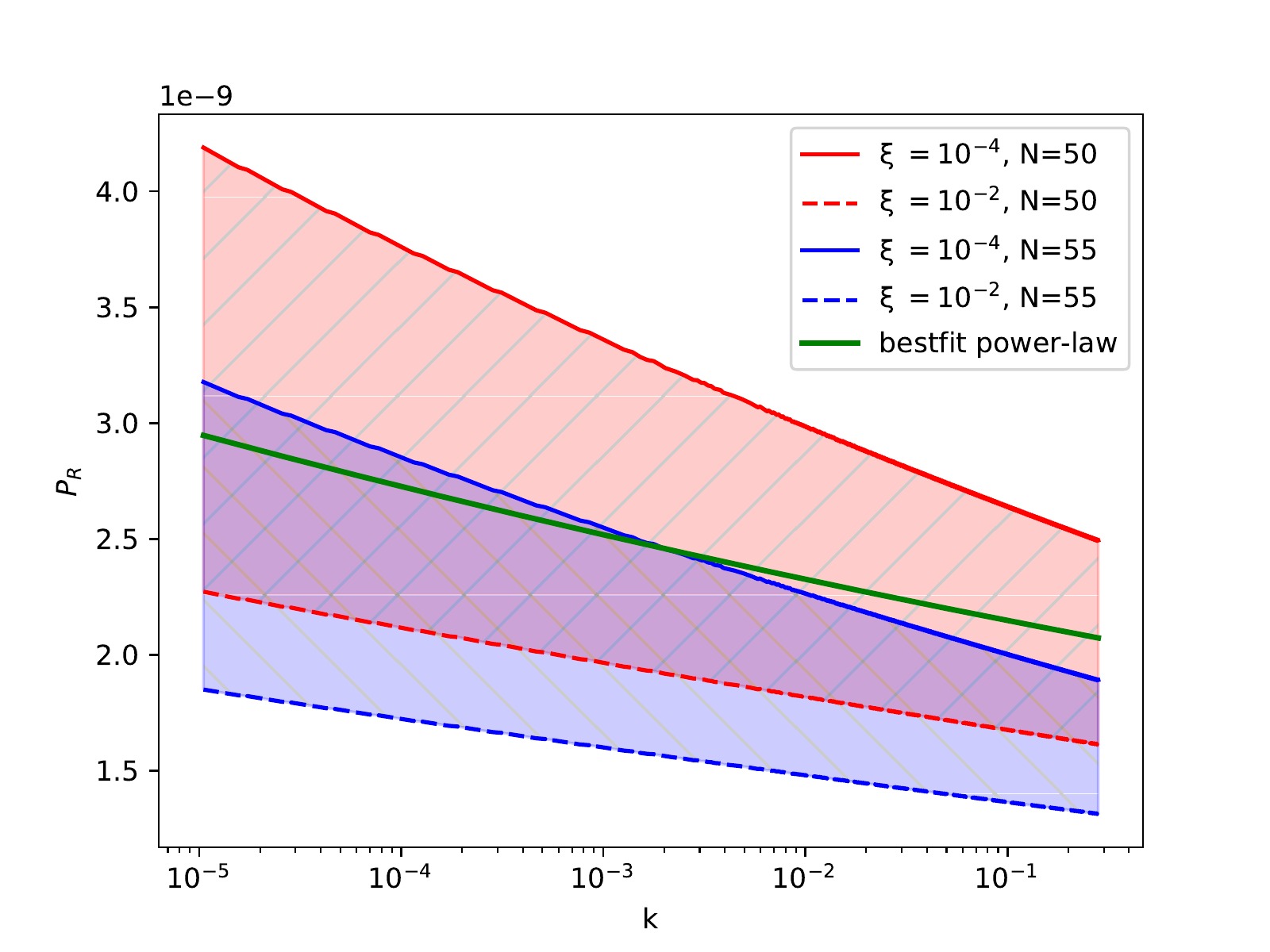}
\caption{{\it Left:} The amplitude $\lambda$ of the CQ potential for different values of the non-minimal 
coupling parameter $\xi$ and different e-folds numbers, $N$. {\it Right:} The Chaotic Quartic primordial power spectrum $P_R$ for different values of the non-minimal coupling parameter $\xi$, assuming $N=50$ (red curves and area) and $N=55$ (blue curves and area).} \label{fig:lambda_xi}
\end{figure*}
The canonical inflationary parameters for the CW potential can be written as
\begin{equation}
 n_s=-\frac{48\xi^2\phi_{*}^2}{(1-2\xi\phi_{*}^2)^2}-\frac{8\xi}{1-2\xi\phi_{*}^2}+1,
\end{equation}
and
\begin{equation}
 r=\frac{128\xi^2\phi_{*}^2}{1-2\xi\phi_{*}^2}.
\end{equation}
A comparison of the theoretical predictions of this model with the $\Lambda$CDM is shown in Fig. (\ref{fig:CW_predictions}), considering the most recent Planck data \cite{Aghanim:2015xee}. In the left panel we show the predicted temperature anisotropy power spectrum for some selected values of $\xi$ whereas in the right panel the $r -n_s$ plane is shown considering two different numbers of e-folds, i.e., $N=50$ and $N=60$ and different values of coupling $\xi$ (represented by the red dots on the green curve). We note that for larger values of $\xi$, $n_s$ rapidly decreases while $r$ increases, which makes the theoretical predictions of the model incompatible with the observational constraints established by standard cosmological model and the joint Planck and Bicep$/$Keck Array data~\cite{Ade:2015tva}.

\subsection{Chaotic quartic (CQ) potential}

The non-minimally coupled chaotic quartic inflation with a spontaneous symmetry breaking was analysed by Linde  et al. in Ref.~\cite{Linde:2011nh}.
The dynamics of this model is governed by the potential
\begin{equation} \label{Linder}
V(\phi)= \frac{\lambda}{4}\phi^4\;,
\end{equation}
which in the Einstein frame it is written as %
\begin{equation}
 V(\phi)= \lambda\frac{\phi^4}{4\Omega^4}\;.
 \label{eq:Potential_phi4}
\end{equation}

The slow-roll parameters are given by
\begin{equation}
 \epsilon= \frac{8\phi^2}{\phi^4(1+\xi\phi^2+6\xi^2\phi^2)}\;,
\end{equation}
\begin{equation}
 \eta= -\frac{48\xi^3\phi^6+24\xi^2\phi^6-48\xi^2\phi^4-4\xi\phi^4-12\phi^2}{\phi^4(1+\xi\phi^2+6\xi^2\phi^2)^2}\;,
\end{equation}
and the number of e-folds can be written as
\begin{equation}
 N = \frac{3}{4}\log\frac{1+\xi\phi_{end}^2}{1+\xi\phi_{*}^{2}}+\frac{1}{8}\left[(1+6\xi)(\phi_{*}^2-\phi_{end}^2)\right]\;,
\end{equation}
where
\begin{equation}
 \phi_{end} =  \sqrt{ \frac{ -1+\sqrt{1+32\xi+192\xi^2} }   {2\xi(1+6\xi)}  }.
\end{equation}
The spectral index and tensor-to-scalar ratio are
 \begin{equation}
 \begin{aligned}
  n_s = &\frac{48}{\phi_{*}^2(1+\xi\phi_{*}^2+6\xi^2\phi_{*}^2)}   \nonumber \\
 & -8\frac{(12\xi^3\phi_{*}^6+2\xi^2\phi_{*}^6-12\xi^2\phi_{*}^4-\xi\phi_{*}^4-3\phi_{*}^2)+1}{\phi_{*}^4(1+\xi\phi_{*}^2+6\xi^2\phi_{*}^2)^2}  \;,
 \end{aligned}
\end{equation}
\begin{equation}
 r= \frac{128}{\phi_{*}^2(1+\xi\phi_{*}^2+6\xi^2\phi_{*}^2)}.
\end{equation}
Similarly to the CW case, we use the COBE normalization  to find the amplitude of the primordial potential in terms of the coupling constant $\xi$, i.e.,
\begin{equation}
 \lambda= \frac{0.16896 \times 10^{-05} \pi^2(1+2\xi\phi_{*}^2+\xi^2\phi_{*}^4)}{\phi_{*}^6(1+\xi\phi_{*}^2+6\xi^2\phi_{*}^2)}.
\end{equation}
The left panel of Fig.~(\ref{fig:lambda_xi}) shows the $\lambda - \xi$ plane for different values of the number of e-folds $N$. Clearly,  the behaviour of $\lambda$ is rather different from the one predicted by Eq. (\ref{lcw}) for the CW model, with the amplitude being almost constant up to values of $\xi \sim 0.0001$. The primordial power spectrum $P_R$ for some selected values of the non-minimal coupling parameter $\xi$, assuming $N=50$ (red) and $N=55$ (blue), is shown in the right panel of Fig.~(\ref{fig:lambda_xi}). As one may see, the Planck power-law best-fit curve (green line) can be completely recovered for values of  $N=50$. We, therefore, assume this value in the statistical analysis presented in the next section. 

\begin{figure*} 
\includegraphics[height=0.35\hsize]{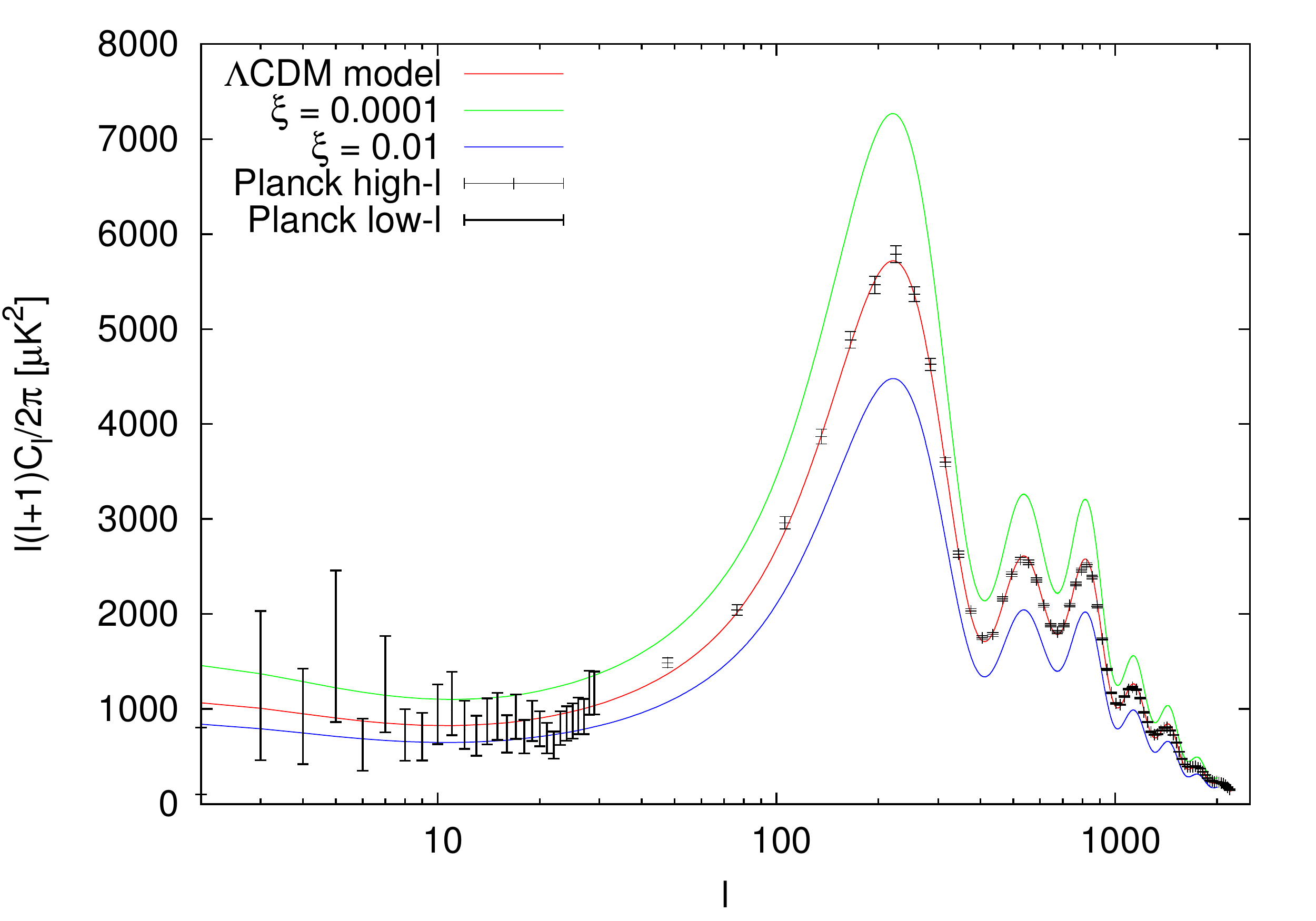}
\includegraphics[height=0.35\hsize]{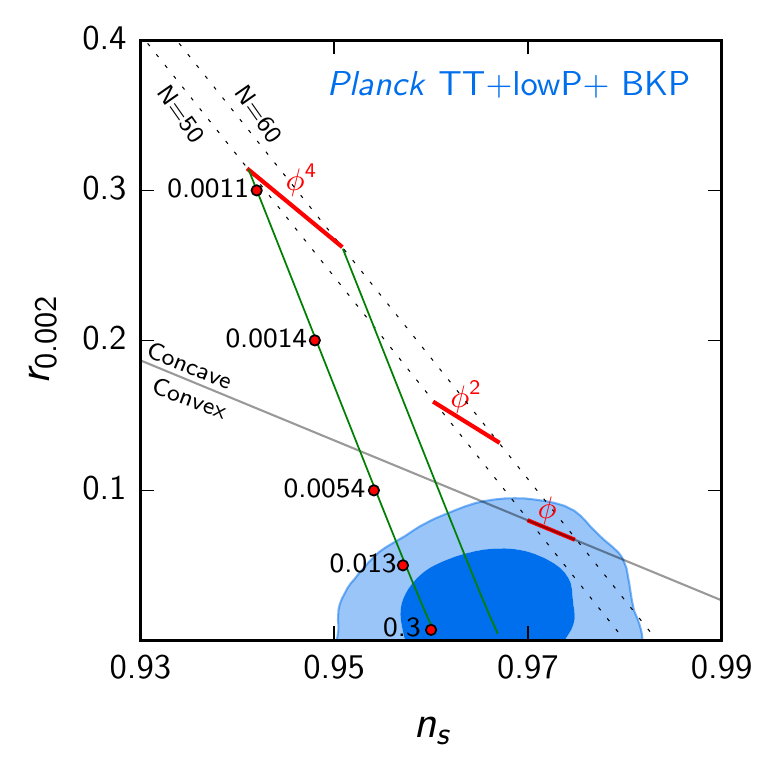}
\caption{ {\it{Left:}} The CMB power spectrum prediction of non-minimally coupled CQ models for selected values of the coupling $\xi$, compared with the most recent Planck data. {\it{Right: }} The $n_s - r$ plane. The contours correspond to $68 \%$ and $95\%$ regions obtained from the Plank TT+lowP data combined with the  Bicep$/$Keck Array data~\cite{Ade:2015tva}. The green lines correspond to the evolution of the $n_s$ and $r$ values for increasing values of $\xi$ (red dots) assuming $N=50$ and $N=60$.} 
\label{fig:phi4_ns_r}
\end{figure*}

Finally, in Fig.~(\ref{fig:phi4_ns_r}) we show the predictions for the temperature anisotropy power spectrum for different values of $\xi$ (left panel), compared with the Planck data. In the right panel we also can see the changes in the  $r-n_s$ plane with respect to the non-minimal coupling constant $\xi$ compared with the Planck+Bicep/Keck Array data.  We see that the $n_s$ value rapidly grows with $\xi$ while the tensor-to-scalar ratio value $r$ drops to zero. The predicted values are inside the $95\%$ C.L. results for values of $\xi \gtrsim 0.012$ (and $N = 50)$. We stress that such a behaviour is the opposite to what happens in the CW case shown in the right panel of Fig.~(\ref{fig:CW_predictions}).
\section{Method and Analysis}
\label{Method}

In order to probe a possible non-minimal coupling in the inflationary phase we perform a Bayesian model comparison considering three models, namely, the standard $\Lambda$CDM scenario (as reference model) and the non-minimally coupled CW and CQ models discussed in the previous section. 

The code most widely used to resolve the Boltzmann equations and explore the cosmological parameter space is the~{\sc CosmoMC} code~\cite{Lewis:2002ah}, in which the theoretical predictions of the models are calculated by  the \textit{Code for Anisotropies in the Microwave Background}({\sc CAMB})~\cite{camb} and the statistical analysis performed with the  Markov-Chain Monte-Carlo method.  We make two main modifications to the most recent {\sc CosmoMC} release.  The first one is in {\sc CAMB}, since it assumes a power-law parametrization for the primordial perturbation spectrum as $P_R=A_s (k/k_*)^{ns-1}$. Here, we use the primordial potential forms shown in Eqs.~(\ref{eq:Potential_CW}) and (\ref{eq:Potential_phi4})  to compute the dynamics and perturbations of these inflationary models and construct their primordial power spectra. We modify {\sc CAMB} following the lines of the~{\sc ModeCode}~\cite{Mortonson:2010er, Easther:2011yq}
adapted for our primordial potentials. This latter is an efficient and high-precision numerical tool able to compute the CMB anisotropies spectrum solving numerically the inflationary model equations. For an exact form of the inflaton potential $V(\phi)$ of a single field model, the code solves the Friedmann and Klein-Gordon equations as well as the Fourier components of the gauge-invariant quantity $u\equiv - z \mathcal{R}$, where $\mathcal{R}$ is the curvature perturbation, $z= a\dot\phi/H$, $a$ is the scale factor and dots denote derivatives with respect to the cosmological time $t$. This set of equations can be integrated to obtain $H$, $\phi$ and thus $z$, as a function of time. From the solution of $u_k$ for the mode $k$, the code can compute the power spectrum of the curvature perturbation $P_\mathcal{R}$ by~\cite{Liddle}
\begin{equation}
P_\mathcal{R} = \frac{k^3}{2\pi^2}\left|\frac{u_k}{z}\right|^2\;.
\label{eq:PRk}
\end{equation}
%
%
\begin{table}[!t]
    \centering
    \begin{tabular}{|c|c|}
        \hline
        Parameter   & Prior \\
        \hline  \hline
        & \\
        $100\,\Omega_b h^2$		& [$0.005$ : $0.1$]  \\
        $\Omega_{c} h^2$			& [$0.001$ : $0.99$]  \\
        $100\, \theta$			& [$0.5$ : $10$]  \\
        $\tau$					& [$0.01$ : $0.8$]  \\
        $100\, \xi_{CW} $		& [$0.0075$ : $0.0085$]  \\  
        $100\, \xi_{CQ} $		& [$0.0001$ : $0.01$]  \\
        & \\
        \hline  
    \end{tabular}
    \caption{\label{tab:priors} Priors on the cosmological parameters considered in the analysis.}
\end{table}
%

The second main modification is made in the {\sc CosmoMC} algorithm, employing the nested sampling of the code {\sc MultiNest}~\cite{Feroz:2008xx,Feroz:2007kg,Feroz:2013hea} to perform our Bayesian analysis of the models. The code {\sc MultiNest} is able to accurately analyse models with non-gaussian density distributions and pronounced degeneracies  in  high  dimensions. It also calculates the evidence with  an  associated  error  estimate, allowing a model comparison in which the ``best'' model is the one that achieves the best compromise between quality of fit and predictivity. Indeed, while a model with more free parameters will always fit the data better (or at least as good as) a model with less parameters,  such added complexity ought to be avoided whenever a simpler model provides an adequate description of the observations. The Bayesian model comparison offers a formal way to evaluate whether the extra complexity of a model is supported by the data, preferring the model that describes the data well over a large fraction of their prior volume (we refer the reader to \cite{Trotta:2005ar,thoven2016,Benetti:2016ycg,Benetti:2017gvm,Graef:2017cfy, Heavens:2017hkr} for some recent applications of Bayesian model selection in cosmology). It is worth mentioning that we use here the most accurate Importance Nested Sampling (INS)~\cite{Cameron:2013sm, Feroz:2013hea} instead of the vanilla Nested Sampling (NS), requiring INS Global Log-Evidence error $< 0.1$.

In our analysis, we vary the usual cosmological parameters, namely, the physical baryon density, $\Omega_bh^2$, the physical cold dark matter density, $\Omega_ch^2$, the ratio between the sound horizon and the angular diameter distance at decoupling, $\theta$, the optical depth, $\tau$ and the additional non-minimal coupling parameter $\xi$. 
We also vary the nuisance foreground parameters~\cite{Aghanim:2015xee} and consider purely adiabatic initial conditions. The sum of neutrino masses is fixed to $0.06$ eV, and we limit the analysis to scalar perturbations with $k_*=0.05$ $\rm{Mpc}^{-1}$.  
The assumed parameters prior are reported in Tab.~\ref{tab:priors}, for the choice on the parameter $\xi$ we use the results shown in Fig.~(\ref{fig:CW_predictions}) and Fig~(\ref{fig:phi4_ns_r}) to establish the prior range from theoretical predictions. Finally, we use the second release of Planck data \cite{Aghanim:2015xee} (hereafter TT+lowP), namely, the high-$\ell$ Planck temperature data (in the range of $30< \ell <2508$) from the 100-,143-, and 217-GHz half-mission TT cross-spectra and the low-P data by the joint TT, EE, BB and TE likelihood (in the range of $2< \ell <29$). 
%
%
\begin{table*}[!t]
\centering
\begin{tabular}{|c|c|c|c|}
\hline
Parameter &${\Lambda}$CDM & Coleman-Weinberg & Chaotic Quartic\\
\hline
\hline
$100\,\Omega_b h^2$ 	
& $2.222 \pm 0.022$  		
& $2.154 \pm 0.017$  		
& $2.202 \pm 0.017$			
\\
$\Omega_{c} h^2$	
& $0.1197 \pm 0.0021$ 	
& $0.1298 \pm 0.0013$ 
& $0.1221 \pm 0.0013$
\\
$100\, \theta$ 
& $1.04085 \pm 0.00045$ 	
& $1.03961 \pm 0.00041$ 
& $1.04060 \pm 0.00039 $
\\
$\tau$
& $0.077 \pm 0.018$	
& $0.033 \pm 0.012$	
& $0.061 \pm 0.014$
\\
$n_s$ 
& $0.9655 \pm 0.0062$ 	
& $ - $ 
& $ - $ 	 
\\
$\ln ( 10^{10}A_s )$  \footnotemark[1]
\footnotetext[1]{$k_0 = 0.05\,\Mpc^{-1}$.}
& $3.088 \pm 0.034$ 
& $ - $ 
& $ - $ 
\\
$100\, \xi $ 
& $ - $ 				
& $ 0.8125 \pm 0.0012 $ 				
& $0.250 \pm 0.052$		
\\
\hline
\hline
$H_0 $  \footnotemark[2]
\footnotetext[2]{[km s$^{-1}$ Mpc$^{-1}$]}
& $67.31 \pm 0.95$ 
& $62.99 \pm 0.49$ 
& $66.23 \pm 0.53$
\\
$\Omega_m$ 	
& $ 0.315 \pm 0.013$  		
& $ 0.383 \pm 0.009$  		
& $ 0.330 \pm 0.008$ %
\\
$\Omega_{\Lambda}$ 	
& $ 0.685 \pm 0.013$  		
& $ 0.617 \pm 0.009$  		
& $ 0.670 \pm  0.008$	
\\
\hline
\hline
$\Delta\chi^2_{best}$ 
& $-$ 
& $-29.2$ 
& $-2.2$ 
\\
$\ln \mathcal{B}_{ij}$ 
& $-$ 
& $-26.3$ 
& $-10.9$ 
\\
\hline
\end{tabular}
\caption{$68\%$ confidence limits for the cosmological parameters using the TT+lowP Planck (2015) data.
The $\Delta \chi^2_{best}$ and the $\ln \mathcal{B}_{ij}$ refer to the difference between the non-minimally coupled models and the $\Lambda$CDM analysis.}
\label{tab:Tabel}
\end{table*} 
%
%
\begin{figure}[!t] %
\centering
\includegraphics[height=0.3\hsize]{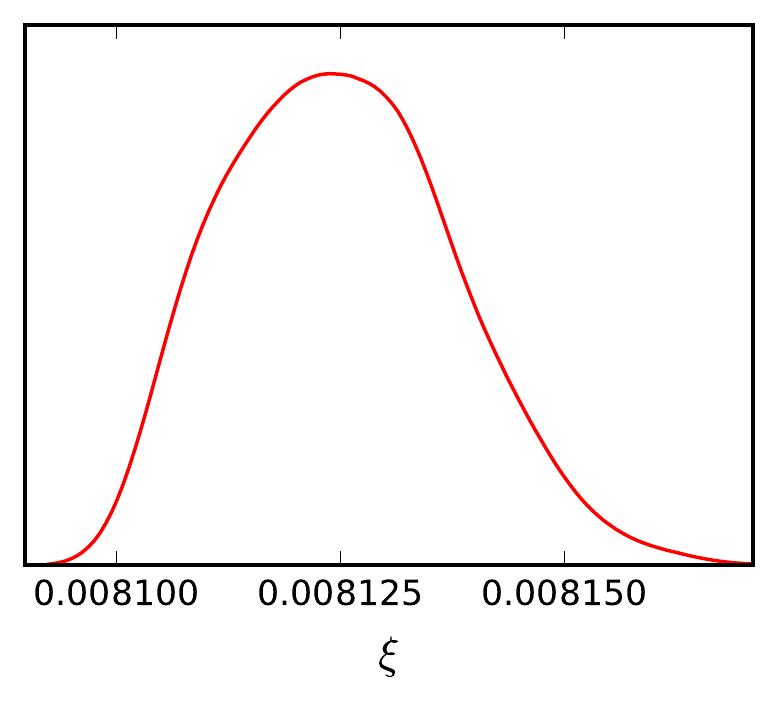}
\includegraphics[height=0.3\hsize]{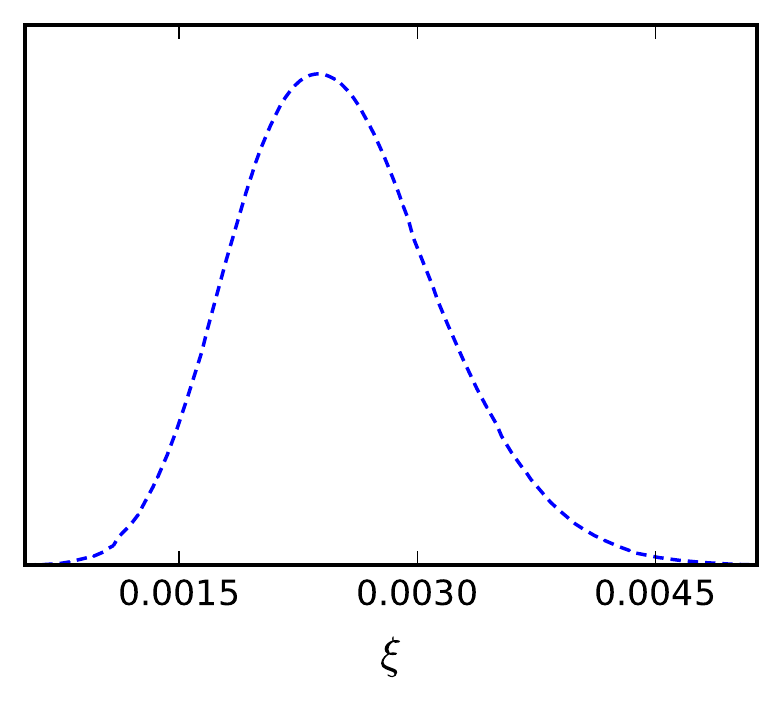}
\caption{  
One-dimensional posterior probability densities for the non-minimal coupling parameter $\xi$ -- see also Tab.~(\ref{tab:Tabel}). The red solid curve corresponds to the CW model (left panel) whereas the dashed blue line refers to the CQ model (right panel).
\label{fig:xi_posteriors}}
\end{figure}
%
\section{Results}
\label{Results}
%
In order to rank the inflationary models discussed in the previous sections, we adopt the Jeffreys' scale to interpret the ratio of the Bayesian evidence, $\ln{\mathcal{B}_{ij}}$, 
 between the model $M_i$ and the reference model ${M_j}$~\cite{Trotta:2005ar, jeffreys}:

\begin{table}[h]
\centering
\begin{tabular}{cccc}
$\ln{\mathcal{B}_{ij}}$ & Odds & Probability &Notes\\
\hline
$< 1$ 	&	$< 3 : 1$ & $< 0.750$ &\text{inconclusive}\\
\nonumber
$1$  & $\sim 3 : 1$ & $0.750$ &\text{weak evidence}\\
\nonumber
$2.5$ & $\sim 12 : 1$ & $ 0.923$ &\text{moderate evidence}\\
\nonumber
$5$    &$\sim 150 : 1$ &$0.993$ &\text{strong evidence}\\
\end{tabular}
\end{table}
\noindent  Note that negative Bayes factor value means support in favor of the reference model $j$.

\begin{figure*}[!t] 
\centering
\includegraphics[height=0.25\hsize]{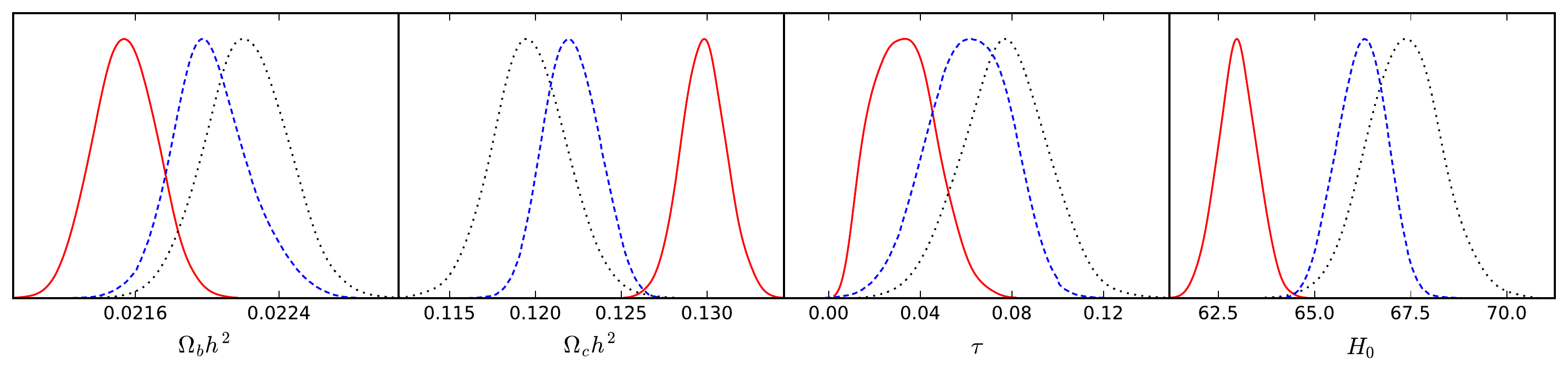}
\caption{ 
One-dimensional posterior probability densities for the analysis shown in Tab.~(\ref{tab:Tabel}). The solid red, dashed blue and black dotted lines  refer to the CW, CQ and $\Lambda$CDM models, respectively.  
\label{fig:cosmol_posteriors}}
\end{figure*}

The main quantitative results of our analysis, i.e., the parameters constraints and the statistical comparison with the $\Lambda$CDM model, are shown in Tab.~\ref{tab:Tabel} while the posterior probability distributions of non-minimal coupling and cosmological parameters are displayed, respectively, in Figs.~(\ref{fig:xi_posteriors}) - (\ref{fig:cosmol_posteriors}). 
We found a very narrow constraint on $\xi$ parameter for both models (see Figs.~(\ref{fig:xi_posteriors})). In the CW case, our result is in disagreement with Ref.~\cite{Panotopoulos:2014hwa}, where the allowed values of $\xi$ were calculated from $n_s$ and $r$ C.L. regions. Instead, we found that the stronger constraint on the non-minimal coupling value is due to its bound with the primordial amplitude.
Also, in Fig.~(\ref{fig:cosmol_posteriors}) we see for the CW model a preference for lower values of $\Omega_bh^2$ and $\tau$ with respect to the $\Lambda$CDM cosmology (black line) and, consequently, for higher values of $\Omega_ch^2$. The tension between the CW and $\Lambda$CDM results can be explained if one considers jointly the tight bound on $\xi$ from the amplitude of the anisotropy power spectrum shown in the left panel of Fig.~(\ref{fig:CW_predictions}) and the influence of this coupling parameter on the $r-n_s$ plane (right panel). Since the values of $r$ and $n_s$ fall outside the confidence contours provided by the Planck+Bicep$/$Keck Array data for the allowed interval of $\xi$, the other cosmological parameter are significantly modified in order to compensate it and fit the data. We also notice the slightly better agreement of the CQ model with the $\Lambda$CDM cosmology. In the CQ model the amplitude of the anisotropy power spectrum also provides the main constraint on $\xi$, with the allowed value of $\xi$ leading to $n_s$ and $r$ values (slightly) out of the data.  

\begin{figure}
\centering
\includegraphics[width=0.8\hsize]{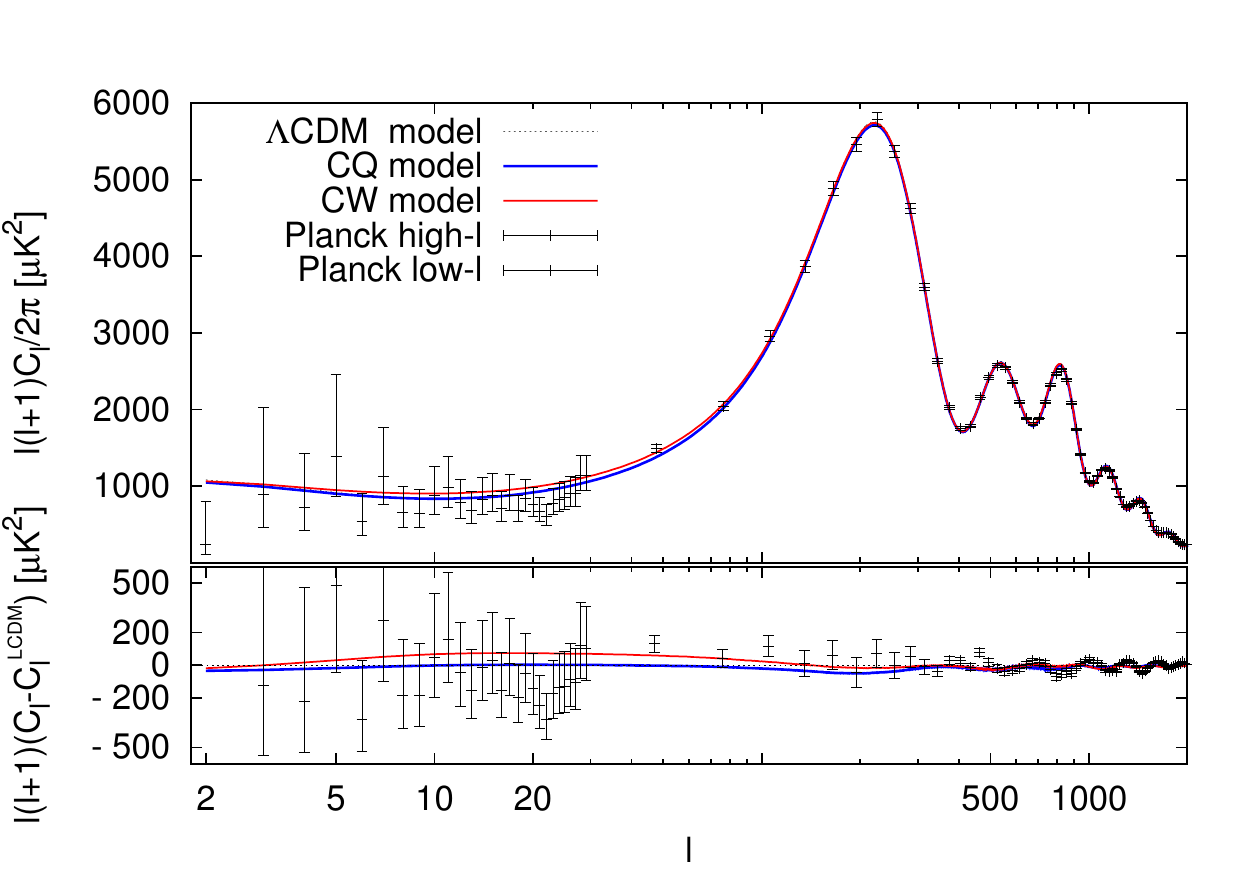}
\caption{Temperature power spectrum for the CW (solid red) and CQ (dashed blue) best fit models. The $\Lambda$CDM best-fit  model (dotted black) is also shown along with the TT+lowP data. In the bottom panel, we show a residual plot with respect to the $\Lambda$CDM prediction.
}
\label{fig:bestfit_TT}
\end{figure}

From the last lines of Tab.~\ref{tab:Tabel}, however, we can see that both the CW and CQ models show a worst $\chi^2$ value with respect to the $\Lambda$CDM model in light of the current CMB data. While for the CW model the difference in the $\chi^2$ value is undeniable, for the CQ model the conclusion may not be so obvious. Indeed, the latter model has one free parameter less than the standard minimal $\Lambda$CDM, which is not taken into account in the $\Delta\chi^2$ value.
As discussed in the previous section, the Bayesian model comparison is the best tool to evaluate the model complexity over the data. The Bayes factor of our analysis are showed in the last line of Tab.~\ref{tab:Tabel} and we can finally conclude that both the non-minimally coupled CW and CQ models are {\textit{strongly}} disfavoured with respect to the standard scenario. For completeness,  we also show in Fig.~(\ref{fig:bestfit_TT}) the best fit curves for the analysed models and the residual plot with respect to the $\Lambda$CDM curve, where the small deviation from the standard model predictions is displayed.


\section{Conclusions}
\label{Conclusions}

Over the past years, motivated by some recent theoretical developments~\cite{Kallosh:2010ug} and by the possibility of implementing chaotic inflation with the Higgs field non-minimally coupled to gravity~\cite{Bezrukov:2007ep,Barvinsky:2008ia,GarciaBellido:2008ab,DeSimone:2008ei,Linde:2011nh},  there has been a renewed interest in inflationary scenarios with a non-minimal coupling $\xi$. 

As is well known, a non-null coupling  may significantly modify the theoretical predictions of the simplest inflationary models and, from the observational side, it is also believed that its introduction may help reconcile these models with CMB data. A clear example is given by the chaotic scenario with potential $\lambda \phi^4$. As shown by several authors, if no coupling is assumed, the model is disfavoured by the data since its prediction of $r$ is too large, but the introduction of $\xi$ is apparently able to reconcile the model predictions with observations~\cite{Okada:2010jf, Okada:2015lia, Bezrukov:2008dt, Linde:2011nh, Komatsu:1999mt}. 

In this paper, we have gone a step further in the analysis of the observational viability of this class of models. Using the most recent CMB data from Planck Collaboration~\cite{Ade:2015xua, Ade:2015lrj}, we have analysed the role of a non-minimal coupling between the inflaton field and the Ricci scalar considering  two selected models of small and large field inflation, namely, the Coleman-Weinberg and the chaotic quartic potentials. 

Firstly, we have shown that the coupling parameter $\xi$ is strongly constrained by the COBE normalization via the amplitude of the primordial power spectrum and not only by the $n_s$-$r$ values, as considered in the previous works~\cite{Panotopoulos:2014hwa, Linde:2011nh, Okada:2010jf}.
Secondly, taking the constrained interval as priors on the value of $\xi$ 
a Bayesian statistical analysis has been performed to compare the predictions of the CW and CQ models with the standard $\Lambda$CDM cosmology and determine if the extra complexity of the non-minimal models is supported by the data.  As shown in Table II, the current CMB data \textit{strongly} prefer the minimal standard model over the non-minimally coupled inflationary scenarios considered. This result clearly shows that the compatibility in the $r - n_s$ plane found in previous analysis is not enough to attest the observational viability of this class of inflationary models. A similar analysis considering a general class of non-minimally coupled inflation with potential $V(\phi) \propto \phi^{n}$ is currently in progress and will appear in a forthcoming communication.


\section{Acknowledgments}
M.C. acknowledges financial support from the 
Funda\c{c}\~ao Carlos Chagas Filho de Amparo \`a Pesquisa do Estado do Rio de Janeiro (FAPERJ).
M.B thanks the financial support of FAPERJ - post-doc Nota 10 fellowship. J. S. A. is supported by Conselho Nacional de Desenvolvimento Cientifico e Tecnol\'ogico (CNPq) and FAPERJ. 
The authors thank the use of COSMOMC and the MULTINEST codes. 

\end{document}